\documentclass[conference,a4paper]{IEEEtran}
\IEEEoverridecommandlockouts

\usepackage{enumitem}%

\PassOptionsToPackage{hyphens}{url}\usepackage{hyperref}%
\hypersetup{hidelinks}
\usepackage[nameinlink, noabbrev, capitalize]{cleveref}%
\usepackage{soul}
\usepackage{graphicx} %
\usepackage[caption=false,font=footnotesize]{subfig}
\usepackage{orcidlink} %
\usepackage{xcolor}
\usepackage{cite}%
\usepackage{booktabs}

\def\BibTeX{{\rm B\kern-.05em{\sc i\kern-.025em b}\kern-.08em
    T\kern-.1667em\lower.7ex\hbox{E}\kern-.125emX}}
\begin{document}

\title{BoilerSketch: A TA-Supervised, Diagram-First GenAI Practice for Structured Diagrams in CS1/Early CS2}

\author{
    \IEEEauthorblockN{%
        Ethan Dickey\IEEEauthorrefmark{1}\orcidlink{0009-0007-3706-5253},
        Vivan Tiwari\IEEEauthorrefmark{1}\orcidlink{0009-0009-5504-3255},
        Anvit Sinha\IEEEauthorrefmark{1}\orcidlink{0009-0008-7538-202X}, and
        Andres Bejarano\IEEEauthorrefmark{1}\orcidlink{0000-0003-2611-2855}
    }%
    \IEEEauthorblockA{\IEEEauthorrefmark{1} \textit{Department of Computer Science}\\
        \textit{Purdue University}\\
        West Lafayette, IN 47907}
}

\maketitle

\begin{abstract}
    This innovative practice full paper presents BoilerSketch, a TA-supervised, diagram-first GenAI practice and tablet interface for providing structured visual explanations in CS1 and early CS2 support settings. Large early computing courses routinely face a support bottleneck during labs and office hours because many student questions are best answered with a diagram rather than additional text, yet most AI tutoring tools remain text-forward and unreliable at producing accurate, pedagogically useful visuals.

    BoilerSketch addresses this gap through a dual-pane interaction model that combines chat with a pen-enabled whiteboard for student sketches and a prompting strategy that constrains the model to generate structured, renderable Mermaid diagrams rather than free-form images. To preserve academic integrity, the system is intentionally scoped to conceptual explanation: it forbids executable code and code-level debugging and uses a human-in-the-loop workflow in which teaching assistants remain accountable supervisors who can monitor sessions and intervene when responses require correction, deeper probing, or escalation to live help.

    We report a 45-minute expert evaluation with 21 instructional staff from a large programming course who used BoilerSketch on representative questions and completed a post-use survey. Two-thirds rated the system at least moderately helpful for conceptual understanding and at least moderately useful for typical support tasks. Staff saw the strongest value in routine diagram-based explanations and noted limits in diagram depth and applicability to more advanced topics. We conclude with practical guidance for adopting supervised, diagram-first GenAI support in early computing courses, emphasizing scope-and-escalation rules, prompt-as-policy guardrails, and reliable structured diagram rendering.
\end{abstract}

\section{Introduction} \label{sec:introduction}
Large CS1 and early CS2 courses routinely face a support bottleneck during labs, recitations, and office hours. Students need timely, individualized help, yet many of their questions are not well served by short textual responses alone. Questions about recursion traces, tree traversals, pointer aliasing, and dynamic programming tables are often easiest to answer with a diagram that makes runtime state or structural relationships visible. At the same time, help-seeking in large courses is difficult to manage, and teaching assistants (TAs) are frequently responsible for absorbing much of this demand \cite{wang2024helpme, riese2021challenges, riese2020teaching, yan2025teaching}.

Computing education research has long argued that learning to program requires students to form an explicit model of how programs execute. Sorva characterizes this challenge in terms of the \emph{notional machine}, arguing that understanding runtime behavior is itself a core learning objective in introductory programming \cite{sorva2013notional}. The program-visualization literature complements this view by showing that diagrams and animations can support novice reasoning about state change, control flow, and data relationships, but that their educational value depends heavily on learner engagement and pedagogical integration \cite{naps2002engagement, sorva2013review}. More recent work continues to show that novices use program visualizations to understand unfamiliar code and to verify their assumptions during comprehension tasks \cite{wu2025visualizations}. This literature suggests that visual explanation is often central to how students build and repair their mental models.

GenAI appears to offer a new way to expand access to individualized support. Recent computing education work has shown that large language models can generate programming exercises and code explanations \cite{sarsa2022automatic}, improve the usefulness of programming error messages \cite{leinonen2023error}, and provide on-demand code explanations inside digital course materials \cite{macneil2023ebook}. Course-level deployments have also shown that AI-based tools can provide pedagogically guarded assistance at scale \cite{liu2024cs50, sinha2024boilertai}. More broadly, the field has begun to move from initial reaction toward questions of pedagogical redesign, including prompt-writing activities \cite{denny2024prompt}, student-generated analogies for threshold concepts such as recursion \cite{bernstein2024nesting}, and interactive LLM-based support for novice debugging and code generation \cite{hellas2023helprequests, yang2024debugging, yeh2025bridging}. This shift reflects a broader consensus in the field: the key question is not whether GenAI will affect computing education, but how to integrate it into educational practice in ways that preserve learning, equity, and instructional judgment \cite{lau2023ban, denny2024computing, bull2024generative}.

However, most prior work on AI support in computing education remains text-first or code-first. Existing systems typically generate exercises, explain code, improve error messages, or support debugging through dialogue. By contrast, classic program-visualization systems provide strong visual representations of execution, but they are generally not conversational, multimodal, or embedded in a supervised help workflow \cite{sorva2013review}. Recent work on augmenting human TAs with AI feedback further suggests that hybrid approaches can be promising but still fragile, especially when human review becomes perfunctory or overreliant on model output \cite{ahmed2025feasibility}. Comparatively little work has explored a diagram-first form of AI-supported help that is explicitly constrained to conceptual explanation and deliberately situated under TA oversight. This leaves a gap between two strong strands of prior work: visualization systems that show program behavior, and AI systems that talk about it.

This paper addresses that gap with BoilerSketch, a TA-supervised, diagram-first GenAI practice for CS1/early CS2 support. BoilerSketch combines a dual-pane tablet interface, student sketches, a prompt that constrains the model to conceptual help and structured Mermaid diagrams, and a human-in-the-loop workflow in which TAs remain accountable supervisors. Rather than treating AI as an autonomous tutor, BoilerSketch uses AI to generate inspectable visual explanations within a bounded instructional workflow.

This full innovative-practice paper makes three contributions. First, it presents a replicable instructional practice for supervised, diagram-centric AI support in early computing courses. Second, it articulates a deployable design pattern in which prompt-level guardrails, structured diagram rendering, and TA escalation rules jointly bound AI assistance around conceptual explanation rather than answer production. Third, it reports an expert evaluation with instructional staff and uses those results to identify both best-fit uses and boundary conditions for adoption.

\section{Connection to Literature} \label{sec:related}
\subsection{Visual Representations, Notional Machines, and Program Visualization}
A long line of computing education research argues that novice programmers struggle not only with syntax and problem decomposition, but also with understanding the runtime behavior of programs. Sorva synthesizes this challenge through the notion of the notional machine, arguing that students must construct an explicit model of how a program executes and how data and control evolve over time \cite{sorva2013notional}. This perspective is especially relevant in early courses, where misconceptions about state, references, recursion, and control flow often persist even when students can produce superficially correct code.

Program visualization research offers one response to this challenge. Naps et al. argue that visualization technologies are educationally valuable only when they engage learners actively rather than positioning them as passive viewers \cite{naps2002engagement}. Sorva, Karavirta, and Malmi's review of generic program-visualization systems similarly concludes that such tools can support introductory programming, but that their effectiveness depends on how well they are embedded into pedagogy and how learners interact with them \cite{sorva2013review}. In other words, what matters is not just whether students see a diagram, but whether that diagram becomes part of a meaningful explanatory exchange.

Recent work continues to reinforce this interpretation. Wu et al. show that novices use program visualizations to understand unfamiliar programs, verify assumptions, and build confidence during comprehension tasks \cite{wu2025visualizations}. BoilerSketch draws on that tradition instead of trying to replace it, treating visual explanation as a first-class pedagogical resource inside a conversational, supervised workflow.

\subsection{Generative AI in Computing Education}
The current literature on GenAI in computing education has rapidly expanded from capability demonstrations to curricular and pedagogical questions. Denny et al. synthesize this shift at the field level, arguing that the advent of code-generating models requires educators to rethink both what students should learn and how they should learn it \cite{denny2024computing}. Lau and Guo's instructor interview study captures the same moment from the perspective of programming instructors, showing that early responses to AI tools often centered on cheating concerns, but that longer-term views increasingly turned toward integration into curriculum and practice \cite{lau2023ban}. Bull and Kharrufa make a similar argument from software development education, advocating deliberate pedagogical integration rather than instinctive prohibition \cite{bull2024generative}.

Within computing education specifically, several strands of work are especially relevant to BoilerSketch. Instructor-facing work has shown that large language models can generate programming exercises and code explanations that are often useful with human review \cite{sarsa2022automatic}. Student-facing work has used large language models to make programming error messages more interpretable and actionable for novices \cite{leinonen2023error}, and to provide multiple styles of code explanations embedded directly in course materials \cite{macneil2023ebook}. Other work treats LLMs not only as support tools but also as objects of instruction: Prompt Problems ask students to learn prompt design as a new programming-adjacent skill \cite{denny2024prompt}, while Bernstein et al. show that students can use LLMs to generate personally meaningful analogies for difficult concepts such as recursion \cite{bernstein2024nesting}.

A second relevant strand focuses on interactive and help-seeking uses of LLMs. Hellas et al. study how LLMs respond to beginner programmers' help requests and show that the models are often useful but still unreliable, with frequent false positives and a tendency to provide model solutions even when prompted not to \cite{hellas2023helprequests}. Yang et al. examine novice debugging with an AI tutor and show that the value of the system depends on when students seek help and how they interpret its responses \cite{yang2024debugging}. Yeh et al. further show that interactive LLM support can improve novice code-generation performance and even improve later prompt construction \cite{yeh2025bridging}. These studies strongly motivate pedagogically bounded interaction. They also make clear that design details matter: support is not simply about access to a model, but about how the interaction is structured. One related framework makes this point explicitly: Dickey et al.'s AI-Lab positions GenAI use in programming courses as a scaffolded, deliberately guided activity intended to preserve core skill development while helping students learn to use the tool productively \cite{dickey2024ailab}.

BoilerSketch builds on this GenAI literature while differing from most of it in an important way. Most existing systems are text-forward. They explain code, rewrite messages, answer questions, or scaffold debugging through language. BoilerSketch instead asks what happens when the primary explanatory artifact is a structured, inspectable diagram generated on demand in response to a student's question or sketch.

\subsection{Help-Seeking, TAs, and Supervised Support in Large Courses}
BoilerSketch also sits within literature on large-course support and TA-mediated instruction. Help-seeking studies show that office hours and other out-of-class support spaces are valuable but difficult to scale. Wang and Lawrence document recurring logistical and pedagogical challenges in managing office-hours participation and demonstrate that support systems can reshape student engagement with help channels \cite{wang2024helpme}. For our purposes, this literature matters because it frames support not simply as a technical delivery problem, but as an interactional and organizational challenge within large courses.

In computing education, TAs are central to handling this challenge. Riese and Kann show that tutoring and assessment are substantial and complex parts of the TA role in computer science education \cite{riese2020teaching}. Riese et al. extend this perspective across institutions and countries, identifying recurring challenges related to professional identity, student interaction, assessment, and best practice \cite{riese2021challenges}. Subsequent work on TA preparation emphasizes that these roles should be intentionally scaffolded rather than treated as informal labor \cite{riese2022training, yan2025teaching}. This literature suggests that any system intended to operate in labs or office hours must be designed around TA practice instead of displacing it.

This point is especially salient in light of recent work on AI-augmented TAs. Ahmed et al. report mixed results when human TAs were provided AI-generated feedback for CS1 programming exercises, noting both perceived benefits and risks of complacency and over-reliance \cite{ahmed2025feasibility}. That finding aligns closely with our design stance. BoilerSketch keeps TAs in the loop and preserves their instructional authority. The AI generates a first-pass artifact, and the TA inspects, extends, or rejects it.

\subsection{Positioning the Present Work}
The present paper connects these literatures by treating AI support as an extension of instructional practice rather than as a replacement for instruction. Unlike classic program-visualization systems, BoilerSketch is conversational and multimodal. Unlike most LLM-based computing-education tools, it is diagram-first rather than text-first, and it explicitly refuses code generation in favor of conceptual explanation. Unlike autonomous AI tutor framings, it places TAs at the center of the workflow as accountable supervisors.

The contribution is therefore the combination: structured visual representation, prompt-level pedagogical guardrails, and TA-mediated deployment for early-course support. That combination, we argue, is the literature-grounded reason to study BoilerSketch as an innovative practice rather than as a generic tutoring system.

\section{Innovative Practice: TA-Supervised Diagram-First GenAI Support}

\subsection{Practice Overview}
Large CS1/early CS2 courses generate many help requests for which the immediate instructional need is not a full solution, but a fast and inspectable explanation. Many of these requests are diagram-centric: students need to see how a data structure changes over time, how an algorithm traverses a graph or tree, or how state evolves across a trace. BoilerSketch was designed for this specific niche. Rather than treating GenAI as a general tutor, the practice scopes AI to first-pass conceptual explanation, especially when a visual representation can make sequence, state, or relationships easier to understand.

The intended setting is a staffed learning environment such as lab, recitation, or office hours. Students interact with a tablet-based interface that supports both text and sketches. When a question appears to benefit from visualization, the system prompts the model to produce a structured diagram specification, renders that specification into a viewable diagram, and returns the diagram with a brief explanation. Human instructional staff remain accountable for the interaction. In the intended deployment, a teaching assistant can monitor several ongoing sessions and intervene when an answer requires correction, elaboration, or escalation to live help.

This configuration makes BoilerSketch an instructional practice rather than a standalone software artifact. The key contribution lies in how those diagrams are produced. They come with explicit pedagogical constraints, a learner-facing workflow, and a supervision model that preserves instructional control.

\subsection{Core Design Principles}
BoilerSketch is organized around three design principles.
First, the system supports conceptual explanation rather than answer production. BoilerSketch is intended for questions that benefit from visualization and conceptual unpacking. It is not intended to provide executable solutions, complete code, or code-level debugging. This boundary is pedagogically important in early computing courses, where a large portion of student help requests involve understanding rather than answer retrieval.

Second, BoilerSketch treats visual explanation as a structured-output problem. Instead of asking the model to generate free-form images, the system constrains visual output to a diagram language that can be rendered predictably and inspected quickly by instructional staff. This choice reflects the practical reality that instructional support tools must produce artifacts that are not only expressive, but also stable enough for routine use.

Third, BoilerSketch keeps instructional authority with humans. The AI response is a first response. Teaching assistants remain responsible for deciding when a response is sufficient, when a misconception needs to be addressed more directly, and when the interaction should shift from AI-supported explanation to live human assistance. This principle is central to the practice because many early-course questions appear straightforward on the surface while concealing deeper confusion.

Taken together, these principles define the scope of BoilerSketch more clearly than any individual interface feature. The practice is deliberately narrow: it aims to provide rapid, inspectable, concept-level visual explanations under human supervision.

\subsection{Workflow and Learner Interaction}
\begin{figure}[t]
    \centering
    \subfloat[Text-only request with returned max-flow/min-cut diagram in the chat interface.\label{fig:bs_maxflow_chat}]{
        \includegraphics[width=0.97\columnwidth]{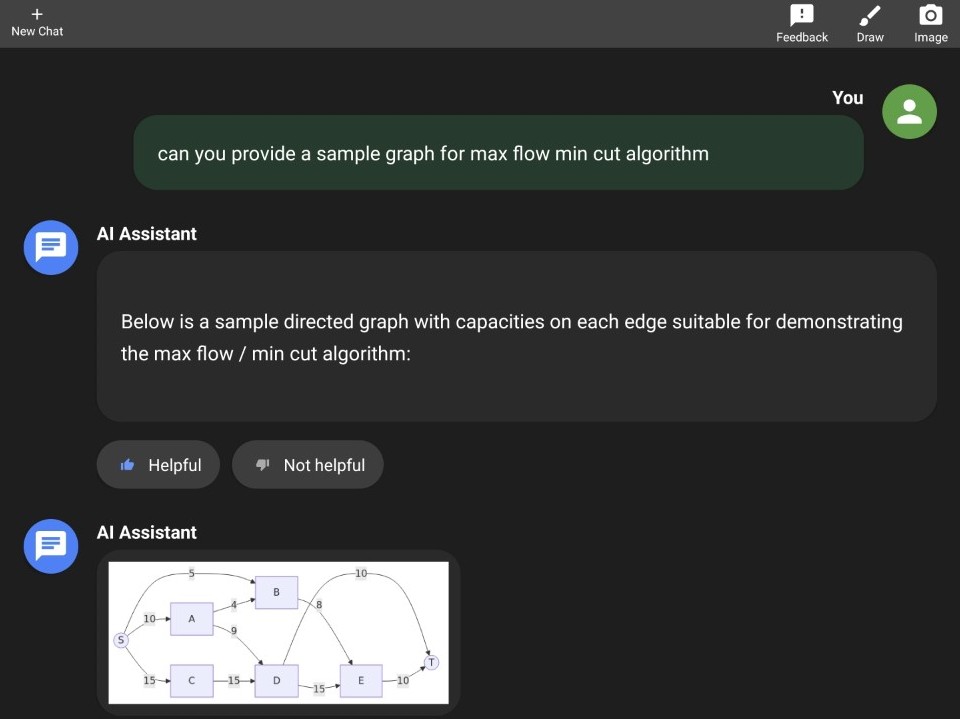}
    }
    
    \vspace{0.4em}
    
    \subfloat[Student-provided hand-drawn red-black tree used as context.\label{fig:bs_rbtree_input}]{%
        \includegraphics[width=0.48\columnwidth,height=0.28\textheight,keepaspectratio]{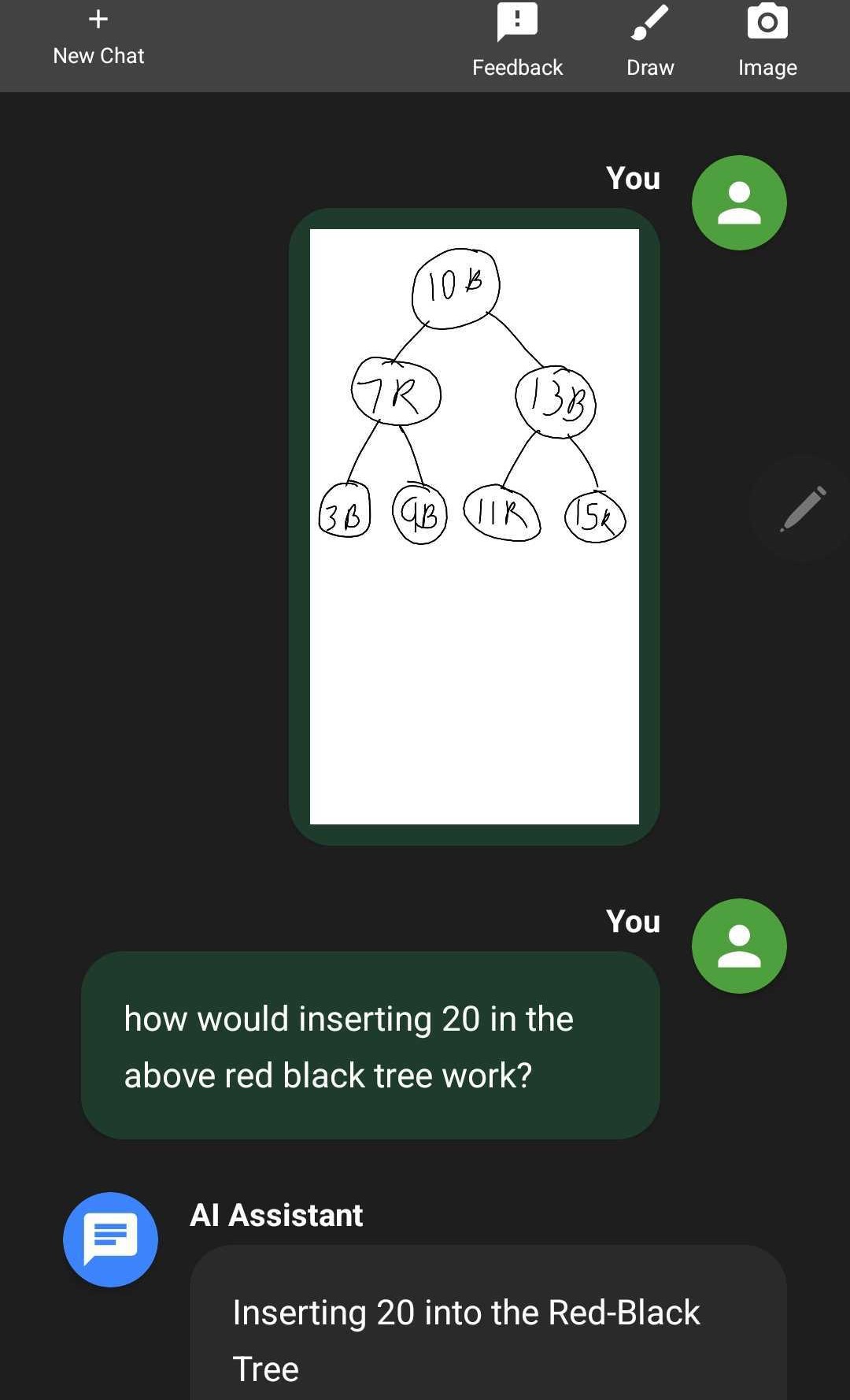}
    }
    \hspace{0.01\columnwidth}
    \subfloat[Returned structured tree visualization after the follow-up explanation.\label{fig:bs_rbtree_output}]{
        \includegraphics[width=0.48\columnwidth,height=0.28\textheight,keepaspectratio]{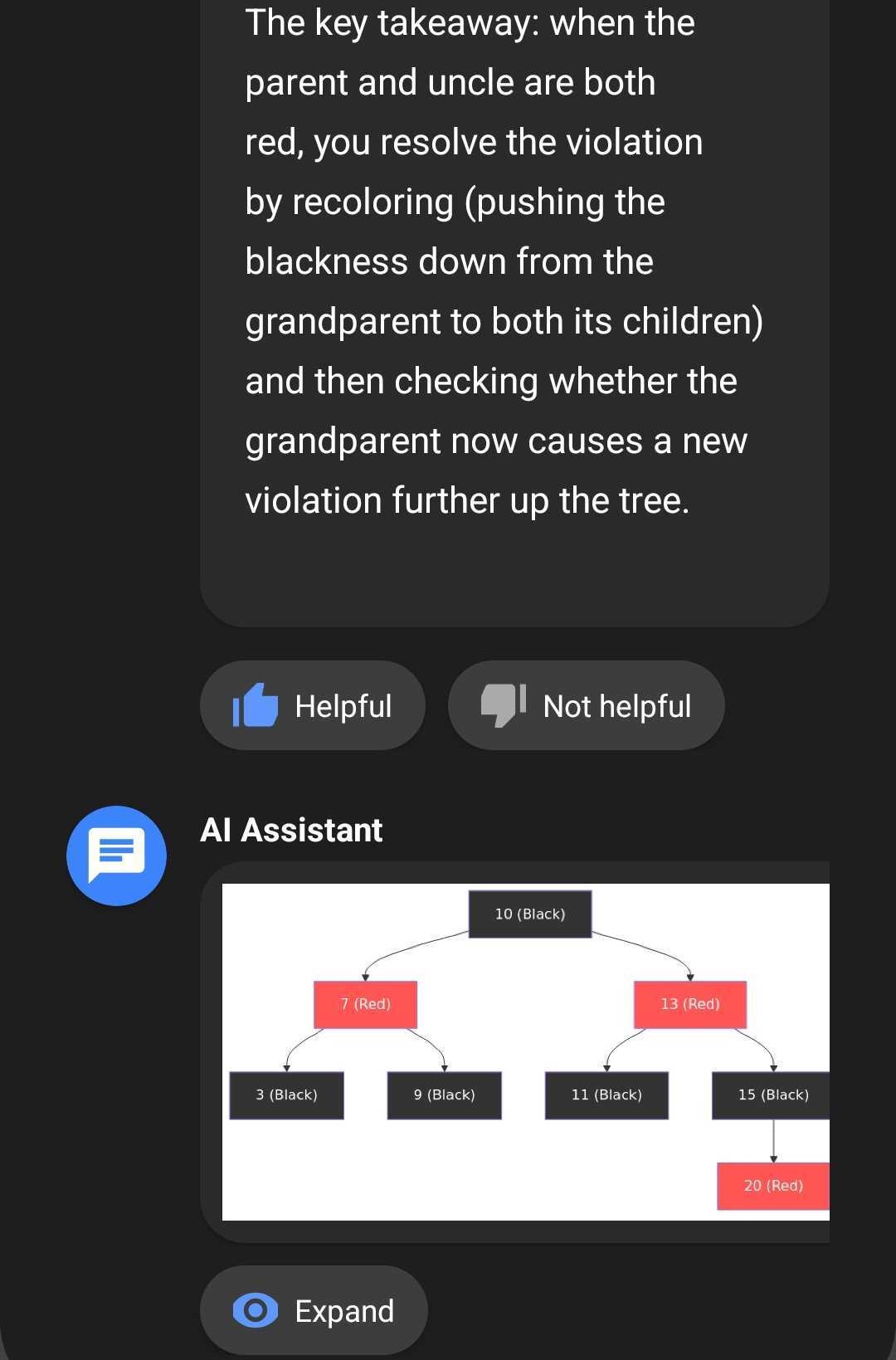}
    }
    \caption{Representative BoilerSketch interactions. The top panel shows a text-only request that produces a rendered max-flow/min-cut diagram in the chat interface. The bottom row shows a sketch-anchored interaction in which a student-provided red-black tree is used as context for a follow-up conceptual question and a returned visualization. Together, these examples illustrate both text-to-diagram and sketch-anchored support workflows.}
    \label{fig:boilersketch_screens}
\end{figure}

BoilerSketch uses a dual-pane tablet interface that supports both chat and sketching. In chat view, students enter natural-language questions and receive a short explanation together with a rendered diagram when a visual response is warranted. In whiteboard view, students can sketch a structure, trace, or partial solution, then attach that sketch to a question. Students may also take a picture of a hand-drawn diagram (e.g., from a whiteboard discussion with a TA) and attach that to a question. Returned diagrams can be enlarged, saved, and used as the basis for follow-up questions or annotation. This supports iterative visual-textual dialogue rather than one-shot prompting. Representative BoilerSketch interactions are shown in \Cref{fig:boilersketch_screens}. The top panel illustrates a text-to-diagram exchange for a max-flow/min-cut query, while the bottom row shows how a student-provided sketch can be used as context for a follow-up conceptual explanation and returned visualization. Additional screenshots of extended multi-turn interactions are provided in the supplementary material\footnote{Supplementary material: \href{https://docs.google.com/document/d/e/2PACX-1vQC6UMPKRU5egA2LlLOpcONc7lrd3QB283cJAFMknOYv_IdQmi6-ezdceEkmdPnAbf5zGFoZksAEiTa/pub}{extended BoilerSketch interactions.} \href{https://github.com/etdickey/BoilerSketch}{Project repository}.}.

The underlying workflow is straightforward. A student begins in chat view, optionally moves to the whiteboard to draw or annotate a sketch, and submits the question to the server. The model then produces a response under the system prompt described below. If the response includes a diagram block, the server extracts the structured diagram description, renders it into an image, and returns the diagram together with any accompanying explanatory text.%

The educational value of this workflow lies in its support for bidirectional visual communication. Students can contribute a sketch when text alone is insufficient, and the system can respond with a visual artifact that can be discussed, revised, or extended across later turns. In early computing courses, this is a closer match to how help is often provided in person, where instructors and students jointly use diagrams, traces, and quick sketches to make reasoning visible.

\subsection{Prompt as Policy}
In BoilerSketch, the system prompt functions as a pedagogical policy artifact. It defines what kind of help the system is allowed to provide and, just as importantly, what kind of help it must refuse. The workflow described above depended on a two-layer prompting strategy. A base instructional prompt governed the assistant's role, tone, and academic-integrity boundaries, while a separate visualization subprompt was invoked only when a visual explanation was warranted. \Cref{tab:prompt_guardrails} summarizes the core guardrails encoded in the base prompt, and the full base prompt together with the visualization subprompt is reproduced in the supplementary material.

\begin{table}[t]
    \caption{Core prompt guardrails used in BoilerSketch}
    \label{tab:prompt_guardrails}
    \centering
    \begin{tabular}{p{0.95\columnwidth}}
    \hline
    1. Use straightforward, professional, and concise language. \\
    2. Focus only on the academic content of the student's question. \\
    3. Prefer guidance that supports independent problem solving; answer direct conceptual questions succinctly only when doing so does not disclose solutions. \\
    4. Do not provide executable code. \\
    5. Do not directly debug or correct student code. \\
    6. When visualization is needed, invoke a Mermaid-only diagram subprompt. \\
    7. Display rendered diagrams to students rather than raw Mermaid source. \\
    \hline
    \end{tabular}
\end{table}

At the instructional layer, the prompt positioned the model as a concise, objective teaching assistant whose role was to support independent problem solving. It directed the model to focus on the academic content of the student's question, to avoid engaging course complaints, and to withhold executable code and direct debugging help. These constraints defined what counted as appropriate assistance before any response was rendered.

When a diagram was likely to clarify state, structure, or process, the system invoked a separate visualization subprompt. The first call produced the explanatory prose; the second produced Mermaid only in a single code block, which was then validated, rendered server-side, and returned alongside the text. This separation allowed BoilerSketch to combine ordinary explanatory text with tightly constrained visual generation while keeping the student-facing interaction focused on computing concepts rather than tool syntax.
The prototype used OpenAI's \texttt{o1} during the Spring 2025 evaluation and mermaid-js/mermaid-cli version 11.4.x.

Our innovative practice is centered around this prompt treatment. It provides a transparent mechanism for translating course values, especially conceptual focus and academic integrity, into system behavior. It also gives instructors a concrete artifact that can be reviewed, revised, and aligned with local course policies. In this sense, prompt design does real pedagogical work. It belongs to the instructional design.

\subsection{Structured Diagram Generation and Rendering}
A key design decision in BoilerSketch is the choice to generate structured diagrams rather than free-form images. During development, unconstrained visual outputs were not reliable enough for instructional use. BoilerSketch therefore treats diagram generation as a constrained representational task. The model emits Mermaid diagram syntax, the server isolates valid diagram blocks, and the diagram is rendered before it is shown in the interface. This pipeline offers a useful middle ground between fully scripted visualization tools and unconstrained generative media. It preserves flexibility across open-ended questions while still producing artifacts that are predictable enough to inspect quickly.

Mermaid is a text-based diagramming syntax whose source definitions are rendered into diagrams \cite{sveidqvist2014mermaid}. It was selected because it can express many visual forms common in early computing courses, including trees, graphs, simple process flows, and table-like state relationships, using lightweight text that language models can generate more reliably than richer graphics specifications. Other diagram languages would likely serve as well. The constraint matters more than the choice of language. It is what makes GenAI usable in instructional settings where text-only responses and free-form images both fall short.

This design also supports iteration. Once a diagram has been rendered, the student can inspect it, ask follow-up questions about it, or annotate it in the whiteboard view and resubmit it as part of the next turn. In effect, the rendered diagram becomes a shared representational object within the interaction.

\subsection{Human Oversight as Part of the Practice}
Human oversight is part of the instructional design for BoilerSketch. The system is intended for staffed support settings in which a TA can monitor several ongoing sessions and step in when a response is incomplete, off-scope, or potentially misleading. In practice, TA intervention can take at least three forms: correcting the content of a response, extending it with a more targeted explanation, or redirecting the student to live support when the issue requires diagnosis of a misconception or course-specific guidance.

This oversight matters because early-course questions often mask deeper confusion. A response that is broadly plausible may still fail to address the particular misconception that motivated the question. BoilerSketch is therefore best understood as a supervised first-response layer. It can handle a portion of routine conceptual explanation, but it relies on human staff to preserve pedagogical quality and to decide when the interaction should escalate beyond AI assistance.

The result is a bounded instructional workflow rather than an autonomous tutor. BoilerSketch can help absorb some routine, diagram-friendly questions, but it does so under explicit human supervision and within clearly defined pedagogical limits.

\section{Expert Evaluation Method}
\subsection{Evaluation Purpose and Questions}
This study evaluates BoilerSketch as an instructional practice. It does not test student learning. The purpose of the evaluation was to determine whether instructional staff judged the practice acceptable and useful for routine conceptual support in CS1/early CS2 settings, and to identify the boundary conditions under which the practice should or should not be deployed.

The evaluation addressed three questions. First, to what extent do instructional staff judge BoilerSketch helpful for conceptual understanding? Second, to what extent do they judge it useful for typical support tasks? Third, what conditions do staff identify as especially appropriate or especially problematic for this form of AI-supported, diagram-centric help?

\subsection{Participants and Setting}
Twenty teaching assistants and one instructional specialist participated in a single expert-evaluation session. The instructional staff included both undergraduate and graduate TAs with direct experience supporting students during labs, office hours, and related help settings. We treat these participants as expert evaluators rather than as stand-ins for students. Because they regularly encounter recurring novice difficulties and course-specific misconceptions, they are well positioned to judge whether a support tool is likely to fit existing instructional workflows.

The evaluation was conducted in Spring 2025 with instructional staff from Purdue University's CS251 (Data Structures and Algorithms), a large undergraduate course, whose routine support responsibilities included lab help, office hours, and conceptual troubleshooting for novice programmers. The evaluated task space was intentionally restricted to diagram-centric conceptual explanations that commonly arise across the CS1 to early-CS2 transition, including tracing runtime state, reasoning about traversals, interpreting pointer or reference relationships, and explaining table-based dynamic-programming structure. We selected this staffing context because these instructors regularly encounter the same novice conceptual breakdowns that motivate BoilerSketch, and are therefore well positioned to judge its fit for early-course support workflows.

\subsection{Materials and Task Space}
Each participant used an Android tablet with BoilerSketch installed. The interface provided two primary input modes: a text chat for natural-language questions and a whiteboard for pen-based sketches; additionally, the interface allowed for arbitrary images to be uploaded, written on, and attached to questions. Participants could alternate freely between these modes, attach a sketch to a query, and continue the interaction over multiple turns.

To anchor the session in plausible support scenarios, participants were given representative conceptual prompts that commonly benefit from diagrams. These prompts were intended to span the kinds of early-course questions that arise in labs and office hours, including questions about recursion traces, traversal order, pointer or state relationships, and table-based reasoning. Participants were also encouraged to pose self-generated questions so that they could test the system against examples drawn from their own instructional experience.

This task design was important for two reasons. It ensured coverage of the diagram-friendly conceptual space that BoilerSketch is meant to support, and it allowed participants to probe edge cases that matter in real instructional work.

\subsection{Procedure}
At the beginning of the session, participants received a brief orientation to the interface and to the intended instructional scope of BoilerSketch. They were then asked to interact with the system as if they were students seeking help. Over a 45-minute hands-on period, participants submitted text questions, attached or created sketches when useful, and iteratively explored the system's responses through follow-up questions and annotations.

We kept the session exploratory so that participants could assess instructional fit across a range of plausible support situations. This allowed participants to test both routine questions and edge cases, and it yielded a mix of planned and self-generated interactions that informed the subsequent results.

The hands-on activity centered on the learner-facing experience. Participants did not run a full live office-hours deployment. They evaluated the practice through direct use of the tool and through their own experience managing similar help interactions. The evaluation therefore captures expert judgment about feasibility and pedagogical fit before any wider student-facing deployment.

\subsection{Measures and Analytic Approach}
Immediately after the hands-on session, participants completed a short post-use survey. Two survey items are central to the present paper: perceived helpfulness for understanding concepts and perceived usefulness for accomplishing typical support tasks. These items used ordinal response categories spanning very/moderately not helpful/useful, a neutral midpoint, and moderately/very helpful/useful. We summarize these responses using counts and percentages.

In addition to the survey ratings, we retained participant comments and a small number of illustrative interactions from the session to contextualize the quantitative findings. We use these qualitative materials descriptively and interpretively. They were not formally coded. Their purpose in this paper is to clarify where staff saw clear value and where they identified limits or failure modes.

This analytic choice matches the scope of the study. The present paper aims to report whether instructional staff saw BoilerSketch as a useful and appropriately scoped form of support, and to explain the conditions under which they judged it credible. It does not attempt to quantify all possible interaction patterns or produce a formal qualitative model of staff perceptions.

\subsection{Analytic Scope and Claims}
This evaluation was designed to assess perceived usefulness, instructional fit, and boundary conditions. It was not designed to measure student learning gains, office-hours throughput, or comprehensive correctness across all topic types. Accordingly, the analysis that follows focuses on descriptive statistics from the post-use survey together with design-oriented interpretation of staff feedback.

Because participants were instructional staff rather than students, the results should be read as an expert evaluation of feasibility and acceptability before any wider student deployment. That limitation is also a strength for the present purpose. Instructional staff are often the first people who can judge whether a new support tool aligns with course norms, common misconceptions, and the practical realities of help provision in large courses. For an innovative-practice paper, that expert judgment is an appropriate basis for evaluating whether the practice is promising, bounded, and transferable.

\section{Results} \label{sec:results}
We read these results as evidence of perceived usefulness and fit. Learning outcomes fall outside what the survey can show. We report the post-use survey descriptively and use participant comments plus representative interaction examples to clarify where instructional staff saw clear value, where they were neutral, and where they identified important limits.

As shown in \Cref{tab:helpfulness}, 14 of 21 participants (66.7\%) rated BoilerSketch at least moderately helpful for understanding concepts, while 4 of 21 (19.0\%) selected neither helpful nor unhelpful and 3 of 21 (14.3\%) selected moderately not helpful. \Cref{tab:usefulness} shows a similar distribution for usefulness in accomplishing typical support tasks: 14 of 21 participants (66.7\%) rated the system at least moderately useful, 3 of 21 (14.3\%) were neutral, and 4 of 21 (19.0\%) rated it moderately not useful.

On each item, 14 of 21 participants responded positively, and no respondent selected the most negative category. Most positive ratings were moderate rather than very positive (11 of 14 for helpfulness; 9 of 14 for usefulness), and seven participants on each item were neutral or moderately negative. The results therefore show majority support alongside substantial reservation. In short, instructional staff saw value in BoilerSketch for common help scenarios while stopping short of treating it as a substitute for direct human support.

\begin{table}[t]
    \centering
    \caption{Instructional staff ratings of helpfulness for conceptual understanding.}
    \label{tab:helpfulness}
    \small
    \begin{tabular}{lrr}
        \hline
        Response option & Count & Percent \\
        \hline
        Very not helpful & 0 & 0.0\%\\
        Moderately not helpful & 3 & 14.29\% \\
        Neither helpful nor unhelpful & 4 & 19.04\% \\
        Moderately helpful & 11 & 52.38\% \\
        Very helpful & 3 & 14.29\% \\
        \hline
        Total & 21 & 100.00\% \\
        \hline
    \end{tabular}
\end{table}

\begin{table}[t]
    \centering
    \caption{Instructional staff ratings of usefulness for typical support tasks.}
    \label{tab:usefulness}
    \small
    \begin{tabular}{lrr}
        \hline
        Response option & Count & Percent \\
        \hline
        Very not useful & 0 & 0.0\%\\
        Moderately not useful & 4 & 19.04\% \\
        Neither useful nor not useful & 3 & 14.29\% \\
        Moderately useful & 9 & 42.86\% \\
        Very useful & 5 & 23.81\% \\
        \hline
        Total & 21 & 100.00\% \\
        \hline
    \end{tabular}
\end{table}

\subsection{Instructional Use Cases} \label{subsec:best-fit-uses}
Session feedback and illustrative interactions observed during the evaluation point to a clear best-fit use case for BoilerSketch: questions where a diagram can externalize state, sequence, or relationships more effectively than text alone. Participants consistently highlighted diagram-centric topics such as tree traversals and dynamic programming tables, and graph-based traces as especially well aligned with the system's strengths. For example, a graph-based interaction that generated a weighted graph and provided a step-by-step visualization of Dijkstra's algorithm was identified as a strong demonstration of the system's explanatory potential.

A second class of interactions suggested value beyond explanation: the system was able to generate novel visual instances (e.g., more complex graph configurations) for practice. These interactions are consistent with the design rationale behind BoilerSketch: the system is most useful when the instructional need is conceptual visualization within a constrained representation, rather than open-ended problem solving.

\subsection{Identified Boundary Conditions} \label{subsec:boundary-conditions}
The evaluation also clarified where the practice should not be oversold.
On each measure, 7 of 21 participants (33.3\%) rated the system as neutral or moderately not helpful/useful, and qualitative feedback helps explain these responses. Some staff expressed concern that an AI answer may appear broadly correct while still missing the specific misconception embedded in a student's phrasing. Others noted that the generated diagrams could lack the depth needed for more advanced topics.

These responses carry as much weight as the positive ratings. They indicate that BoilerSketch is most credible as first-line support for routine, diagram-centric questions and less appropriate as an autonomous tutor for nuanced diagnosis or advanced content. %

\subsection{Scope of the Findings} \label{subsec:what-supported}
Within the scope of this study, the results support three claims. First, instructional staff judged BoilerSketch acceptable and often useful for routine conceptual support in CS1/early CS2 settings. Second, the system's value proposition rests specifically on diagram-centric explanation rather than general AI tutoring. Third, appropriate use depends on maintaining human oversight and explicit escalation to a TA when the question requires misconception diagnosis, adaptive feedback, or deeper pedagogical judgment.

The current evidence does not establish learning gains, comprehensive correctness across topic types, or reduced office hour wait times. Those questions remain outside the scope of the present evaluation.

\section{Discussion and Implications for Practice} \label{sec:discussion}
This paper contributes an evaluation of one specific instructional practice. That practice combines four elements: a diagram-first prompt, structured diagram rendering, a sketch-capable student interface, and TA oversight. Taken together, the results suggest that this configuration can provide a credible first-response layer for routine conceptual questions in CS1/early CS2 while reserving misconception diagnosis, adaptive feedback, and deeper pedagogical judgment for human staff. This interpretation is consistent with recent arguments that GenAI should be integrated into computing education through explicit pedagogical design rather than treated as an autonomous substitute for instruction \cite{bull2024generative, denny2024computing}.

\subsection{Human Supervision and Augmentation} \label{subsec:augmentation-role}
The evaluation points toward augmentation as the right role for BoilerSketch. This conclusion follows both from the generally positive staff ratings and from the concerns participants raised about subtle misconceptions. In early computing courses, many requests for help are routine enough to benefit from a rapid visual explanation, but still require escalation when the student's difficulty is poorly articulated, course-specific, or conceptually deeper than the initial question suggests. The human-in-the-loop model therefore carries real weight in the design. BoilerSketch appears most appropriate when the AI provides an initial explanation and the TA remains responsible for validating, extending, or redirecting that explanation as needed.

\subsection{Structured Diagram Generation} \label{subsec:structured-diagrams}
One of the most transferable insights from BoilerSketch is the decision to treat visual explanation as a structured-output problem rather than an image-generation problem. During prototyping, unconstrained visuals produced limited pedagogical value, which motivated the shift to Mermaid as the system's only permitted diagram language. That decision matters pedagogically as well as technically. A structured diagram language preserves some of the flexibility of open-ended AI assistance while improving renderability, inspectability, and consistency.

For computing educators, the broader lesson is that GenAI need not produce polished images to be instructionally useful. In some cases, constrained visual artifacts that can be rendered reliably and checked quickly by staff may be more educationally valuable than richer but less predictable outputs.
In this study, the payoff of structured output was not that the diagrams were universally complete or sufficient on their own, but that they were often useful enough, and inspectable enough, to support supervised first-pass explanation.

\subsection{Prompt Constraints as Pedagogical Design} \label{subsec:prompt-as-pedagogy}
BoilerSketch also demonstrates that prompt design is not merely a backend engineering task. In this system, the prompt encodes pedagogical boundaries: it encourages independent problem solving, prohibits code disclosure and debugging, and privileges diagrams when a visual representation is likely to clarify the concept. These constraints shape the kind of help students receive and define the division of labor between AI and human staff. What is replicable here is therefore not only the interface, but also the prompt-as-policy approach for translating course values, especially academic integrity and conceptual focus, into system behavior.

A useful implication for adoption is that prompt design should be treated as a policy artifact rather than an implementation detail. It should be reviewed by instructors, aligned with course integrity expectations, and revised when the scope of acceptable help changes.

\subsection{Implications for Adoption} \label{subsec:transferable-lessons}
Three practical lessons emerge from this study. First, scope the system to questions that are both conceptual and diagram-amenable; it should not be presented as a general-purpose tutor. Second, pair structured output with deterministic rendering so that visuals are predictable enough for instructional use. Third, make escalation explicit: students and staff should know when the system's answer is likely sufficient and when a TA should intervene.

These lessons are intentionally modest. They do not imply universal effectiveness nor do they remove the need for human instructional judgment. They do, however, provide a concrete, transferable template for courses that want to experiment with supervised, diagram-first AI support while retaining instructional control.

Although BoilerSketch was developed in a large course, the same supervision pattern may also be useful in smaller programs: an instructor or small staff can oversee routine diagram-amenable interactions and intervene when escalation is needed.

\subsection{Limitations}
This study has several important limitations. First, it is an expert evaluation of perceived fit and usefulness, not a student study and not a test of learning outcomes. Second, the evidence comes from a single exploratory session with a relatively small convenience sample of instructional staff in one institutional context. Third, the evaluation did not measure diagram correctness systematically across prompts or topics, nor did it examine operational outcomes such as office hour throughput, escalation rates, or response quality under live deployment conditions.
Accordingly, the next step is a student-facing deployment in authentic labs or office hours that evaluates diagram correctness, student experience and learning, escalation rates, and support-workflow outcomes.

\section{Conclusion}
BoilerSketch contributes a bounded, replicable instructional practice for supervised, diagram-first GenAI support in CS1 and early CS2 contexts. The practice combines a sketch-capable student interface, prompt-level guardrails that constrain the model to conceptual explanation and structured Mermaid output, deterministic diagram rendering, and explicit TA oversight. In an expert evaluation with 21 instructional staff, two-thirds rated the system at least moderately helpful for conceptual understanding and at least moderately useful for typical support tasks. These findings support BoilerSketch as a credible first-response layer for routine, diagram-amenable questions, but not as a general-purpose tutor or a replacement for human instructional judgment.

The broader takeaway is that AI support in early computing courses becomes more instructionally credible when it is narrowly scoped, structurally constrained, and embedded within clear escalation to human staff. BoilerSketch offers one concrete model of that design stance. Although the present study does not establish student learning gains, correctness across all topic types, or reductions in office-hour wait times, it does show that supervised, structured diagram generation can be a practical and transferable way to expand conceptual support while preserving instructional control.

\section*{Acknowledgment}
This work was funded by Purdue's Innovation Hub (IH-AI-23002) and the Department of Computer Science through the GoBoiler program. OpenAI's ChatGPT (GPT-5.5) was used for sentence-level language editing in the abstract and the body; all content was reviewed and approved by the authors.

\bibliographystyle{IEEEtran}
\bibliography{refs}

\end{document}